\documentclass[aps,prl,twocolumn,showpacs,superscriptaddress,groupedaddress,superscriptaddress]{revtex4-1}  
\usepackage{graphicx}  
\usepackage{dcolumn}   
\usepackage{bm}        
\usepackage{amssymb}   
\usepackage{booktabs}
\usepackage[usenames]{color}
\usepackage{colordvi}
\usepackage{enumitem}
\usepackage{multirow}
\usepackage{calc}

\usepackage[utf8]{inputenc}
\usepackage[intlimits]{amsmath}
\usepackage{epsfig}
\usepackage{mathtools}
\usepackage{rotating}
\usepackage{transparent}

\usepackage{amsfonts}
\usepackage{ulem}

\newcommand{\vivo}{\textit{in vivo }}
\newcommand{\vitro}{\textit{in vitro}}

\newcommand{\Vitro}{\textit{In vitro }}
\newcommand{\etal}{\textit{et al.}}

\newcommand{\rem}[1]{{}}

\def\greenw#1{{}}
\def\bluew#1{{}}

\begin{document}

\title{Motility states in bidirectional cargo transport}

\date{\today}

\author{Sarah Klein}
\affiliation{Laboratory of Theoretical Physics, CNRS (UMR 8627), University Paris-Sud
B\^atiment 210, F-91405 ORSAY Cedex, France}
\affiliation{Fachrichtung Theoretische Physik, Universit\"at des Saarlandes D-66123 Saarbr\"ucken, Germany}
\author{C\'ecile Appert-Rolland}
\affiliation{Laboratory of Theoretical Physics, CNRS (UMR 8627), University Paris-Sud
B\^atiment 210, F-91405 ORSAY Cedex, France}
\author{Ludger Santen}
\affiliation{Fachrichtung Theoretische Physik, Universit\"at des Saarlandes D-66123 Saarbr\"ucken, Germany}

\begin{abstract}
Intracellular cargos which are transported by molecular motors move stochastically 
along cytoskeleton filaments. In particular for bidirectionally transported cargos it is an open 
question whether the characteristics of their motion can result from pure stochastic fluctuations 
or whether some coordination of the motors is needed.
The results of a mean-field model of cargo-motors dynamics, which was proposed by
M\"uller \etal \cite{mueller_k_l2008} suggest the existence of high motility
states which would result from a stochastic 
tug-of-war.
Here we analyze a non-mean field extension of their model,
that takes explicitly the position of each motor into account.
We find that high motility states then disappear. We consider also a mutual motor-motor activation,
as an explicit mechanism of motor coordination. We show 
that the results of the mean-field model are recovered only 
in case of a strong motor-motor activation in the limit of a high number of motors.
\end{abstract}

\maketitle
\section{Introduction}
Almost every cellular function is related to transport processes.
These are necessary in order to maintain concentration 
gradients, but also in order to built and adapt cellular structures. Many of these transport issues are carried out by molecular motors, 
i.e. proteins which perform a directed stochastic motion along the polar filaments of the cytoskeleton. 

The cytoskeleton of eukaryotic cells is composed  of three different kinds of filaments: actin, intermediate filaments and microtubules. 
These filaments give a cell its characteristic shape, but also play a key role in intracellular transport. Especially microtubules (MTs), which have well-defined plus- and minus-ends, are responsible to sustain transport from the nucleus to the periphery and \textit{vice versa}. 

Two families of proteins called molecular motors walk along the MTs - 
kinesin and dynein. The main difference between the two is that kinesin walks from the minus-end, which is in general located 
close to the nucleus, to the plus-end which grows in the vicinity of the cell membrane, while dynein walks in the opposite direction~\cite{alberts2002}. After this processive motion motors detach stochastically 
from the filament and diffuse around before attaching again.

Molecular motors are able to transport very different types of cargo, either individually or by teams of molecular motors, which are 
attached to the same cargo. Transport by teams of molecular motors is particularly relevant for large objects like vesicles or cell 
organelles {\cite{schliwa_2003}. Transporting cargo 
by many motors in a crowded environment is obviously beneficial since it increases the processivity along the MT dramatically 
and also enhances the ability of the motor cargo complex to withstand larger friction forces.}

For several types of cargo, however, a bidirectional  motion along a filament was observed \vivo \cite{Soppina2009} and \vitro \ \cite{Hendricks2010}. These observations suggest that the cargo is transported both by kinesin and dynein motors which are  
attached to the cargo at the same time. The trajectories of these cargos consist of sections of persistent motion and sudden returns. 
One key question is if the observed cargo motion is a result of unknown coordination mechanisms or is driven by 
fluctuations.

Various theoretical models have been suggested, which aim at describing the origin of the complex dynamics of bidirectionally transported cargo \cite{mueller_k_l2008,kunwar2011}. A mean-field model (MF-model) which describes bidirectional cargo motion \cite{mueller_k_l2008} driven by 
two teams of molecular motors, was introduced a few years ago by M\"uller
\etal . They assume that two equally strong teams of molecular motors with
opposed walking directions 
are bound to a cargo at the same time. The model focuses on the force balance between the two motor-teams without taking the motor positions 
explicitly into account. The forces acting on the motors determine their attachment and detachment rates. The cargo's velocity depends on the forces 
acting on the motors and is uniquely determined for a given number of attached motors of "$+$"- and "$-$"-motors. Given the mean-field assumption that forces are equally shared among motors moving in the same direction, the model by M\"uller \etal \ predicts the existence of high motility states in cargo transport 
that may originate from a {purely stochastic} tug of war between oppositely directed motors, rather than being induced by a regulatory mechanism.

In this work we will test this scenario by using a more explicit modeling approach which is inspired by the model recently introduced by Kunwar \etal \ \cite{kunwar2011}. 
In contrast to the MF-model we explicitly consider the motor's positions on the filament, and the couplings between motors and cargo which 
are modeled as linear springs in our model (\underline{e}xplicit \underline{p}osition-\underline{b}ased, EPB-model). As we want here to test the consequences of the tug-of-war mechanism rather than modeling an explicit experimental setup, we consider that "$+$"- and "$-$"-directed motors have the same response to applied forces. Our results show the absence of long-living directed transport states. We 
further extend the model and  introduce a motor-motor activation, which was inspired by experimental observations \cite{welte1998}.

\section{Model description}

In the EPB-model we assume that two teams of motors are tightly bound to a cargo. Each team consists of $N$ "$+$"- and "$-$"-motors, respectively. 

To determine the load force applied on the cargo at position $x_C(t)$ at time $t$ we take every single motor position $x_i$ into account. We model the motor linker as a linear spring with spring constant $\alpha$ and untensioned length $L_0$, such that motors experience no force when located at a distance smaller than $L_0$ from $x_C(t)$. So the force $F_i(x_C(t),\{x_i\})$ on the cargo caused by the $i$-th motor is given by

\begin{flalign}  F_i& (x_i -x_C(t)) = \label{eq_force} \\ \nonumber
& \ \ \ \ \ \begin{cases}
\alpha (x_i-x_C(t) +L_0), \ \ \  \ &  x_i-x_C(t)<-L_0\\
0 , & |x_i-x_C(t)|<L_0\\
\alpha (x_i-x_C(t)-L_0),  &x_i-x_C(t)>L_0.
\end{cases} \nonumber
\end{flalign}
We assume that no force is exerted if the motor is not attached to the filament.
When attached the motors can perform steps along the filament with a rate depending on the force $F_i$
\begin{align}
s_+(F_i)   = 
\begin{cases}
\frac{v_F}{d}  ,&F_i<0 \\
\frac{v_F}{d}\left[1-\frac{F_i}{F_S}  \right] , &0\leq F_i \leq F_S \\
\frac{v_B}{d} , &F_i > F_S 
\end{cases}
\end{align}
for "+"-motors and {symmetrically}
for "$-$"-motors.
If no force is applied on the motors or if the force is in the direction of its motion the motor moves with its force-free velocity $v_F$ divided by the step length $d$. "+/$-$"-motors continue stepping in their preferred direction if the force is smaller/bigger than the stall force $(+/-)F_S$ but the velocity decreases linearly when the force applied to it increases. If the force (in absolute value) exceeds the stall force the motors step backwards with $v_B \ll v_F$.\

{ Please note that we don't consider exclusion of the motors on the filament, since the number of possible binding sites in close proximity of 
the cargo is huge compared to the number of attached motors. } 

Next to the stepping rates also the motors' detachment rates $k_d(F_i)$ are  force-dependent and explicitly given by
\begin{align}
k_d(F_i) = k_d^0\exp\left(\frac{|F_i|}{F_D}\right),
\end{align}
where $k_d^0$ gives the force-free detachment rate and the detachment force $F_D$ gives the force scale. Once a motor is detached it attaches again with a rate $k_a$ within the tensionfree area $x_C(t) \pm L_0$.

\begin{table}[tb]
\begin{tabular}{||c|c||c|c||c|c||} 
\hline
$v_F$ & 1000 nm/s & $F_D$ & 3 pN & $\alpha$ & $0.1 $ pN/nm\\ 
$v_B$ & 6 nm/s & $k_d^0$ & 1 s$^{-1}$ & $L_0$ &  $110$ nm \\ 
$F_S$ & 6 pN & $k_a$ & 5 s$^{-1}$& $d$ &  8 nm \\ 
\hline
\end{tabular}
\caption{Simulation parameter taken from \cite{mueller_k_l2008} and \cite{kunwar2011}.}\label{parameter}
\end{table}

In the MF-model \cite{mueller_k_l2008} the cargo moves with a constant velocity which is determined only by the number of attached motors of each team between two motor events (at-/detachment). To calculate this unique cargo velocity M\"uller \etal \ introduce an artificial force which assures a force balance between the two teams of motors with equal sharing of the force within one team. 
All motors are implicitly assumed to move with the same constant velocity
as the cargo.

In contrast, here we take the individual motor steps into account.
The force applied on the cargo is determined separately for each motor
(Eq.~(\ref{eq_force})) and is modified when the motor hops.
We have thus to define how the cargo reacts on the force changes induced by
motor hopping. We move the cargo of mass $m$ with radius $R$ along its equation
of motion in a viscous medium with viscosity $\eta$ \footnote{$\eta = 10$ mPas,
$m= 10^{-14}$ kg, $R=1000$ nm} 
\begin{align}
m \frac{\partial^2 x_C(t) }{\partial t^2 } = -6 \pi \eta R \frac{\partial x_C(t)}{\partial t} + \sum_{i=1}^{n_++n_-}F_i(x_C(t),\{x_i\}).
\end{align}
Here $n_+$ and $n_-$ describe the "+"- and "$-$"-motors, respectively, which are attached to the filament, such that $0\leq n_+ \leq N_+$ and $0\leq n_- \leq N_-$.\\
In the second part of this paper, we shall consider a possible synchronization of molecular motors via mutual motor activation in a way we describe later.
\section{Results}
In this work we analyze whether a model which takes every single motor position into account produces the same high motility state as it was seen for strong motors in the MF-model \cite{mueller_k_l2008}. 
Therefore, we consider the same kind of strong motors, as defined in
TABLE~\ref{parameter}, and we measure the velocity distribution and the
probability of a given number of attached motors of each kind $p(n_+,n_-)$.

In the MF-model \cite{mueller_k_l2008} the cargo velocity is
simply a function of attached motors
and is thus constant between two attachment/detachment events.
In our case the cargo moves according to its equation of motion and, therefore, not with constant speed. This is why we discretize the trajectory and define
%
$\overline{v}(t) \coloneqq \frac{x_C(t+\Delta t) - x_C(t)}{\Delta t}$.
%
Here we use $\Delta t = 0.16 $ s.
\subsection{No activation }
\begin{figure}[bt]
\begin{minipage}[hbt]{0.25\textwidth}
\begin{center}\textbf{(a)}\end{center}
\includegraphics[width=0.8\textwidth]{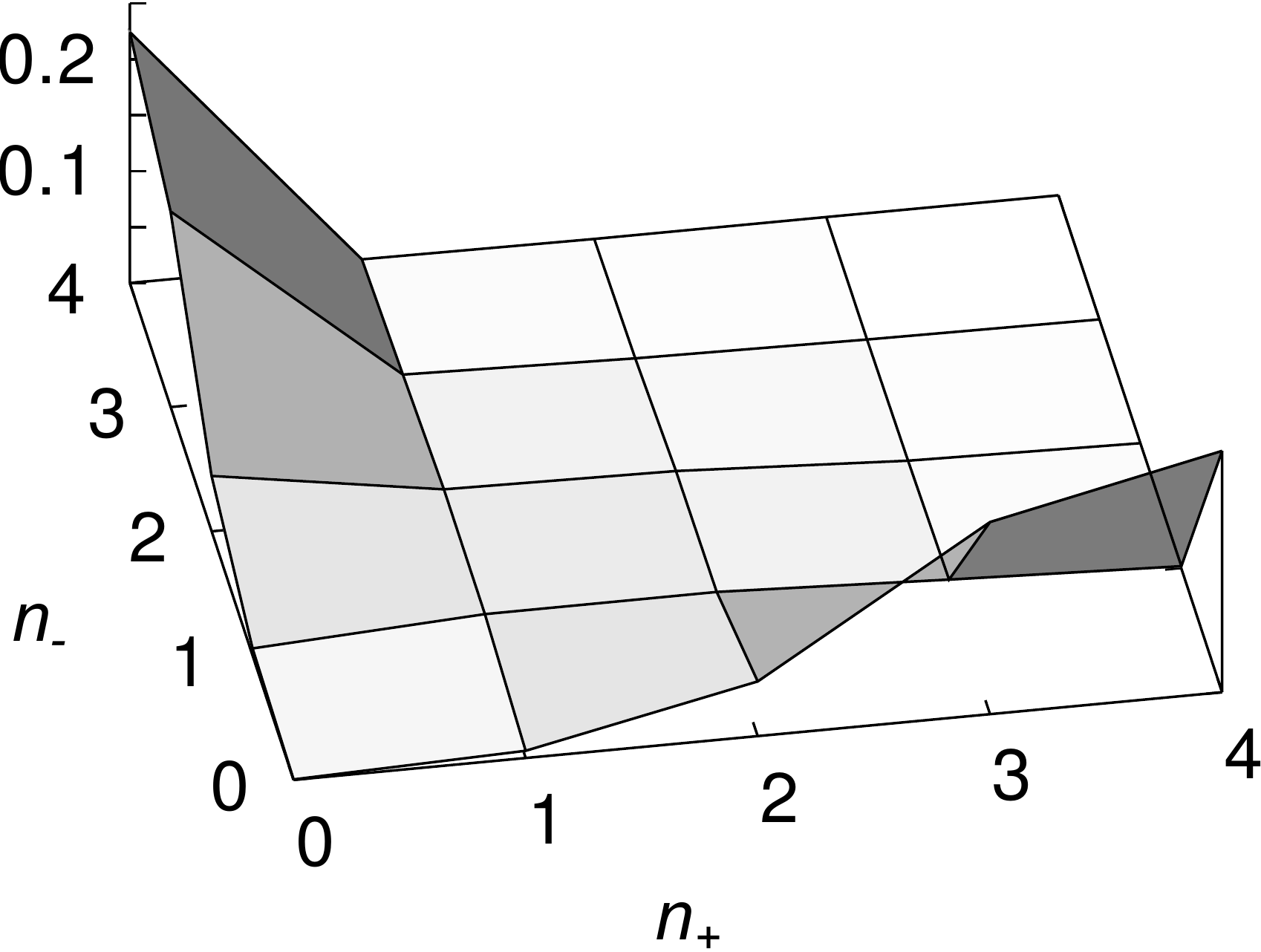}
\end{minipage}\begin{minipage}[hbt]{0.25\textwidth}
\begin{center}\textbf{(b)}\end{center}
\includegraphics[width=0.8\textwidth]{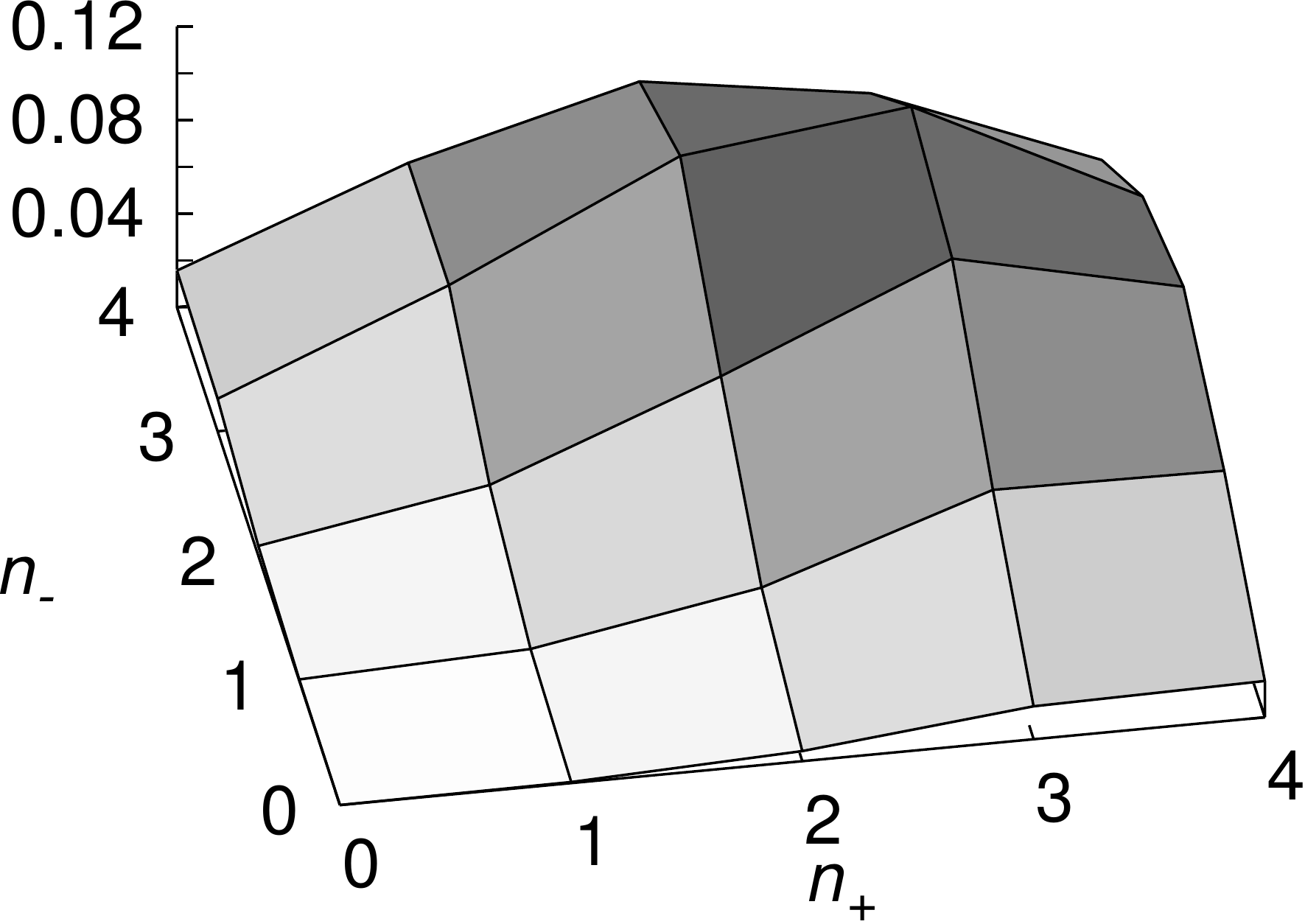}
\end{minipage}\\
\vspace*{0.2cm}
\begin{minipage}[hbt]{0.25\textwidth}
\begin{center}\textbf{(c)}\end{center}
\includegraphics[width=0.8\textwidth]{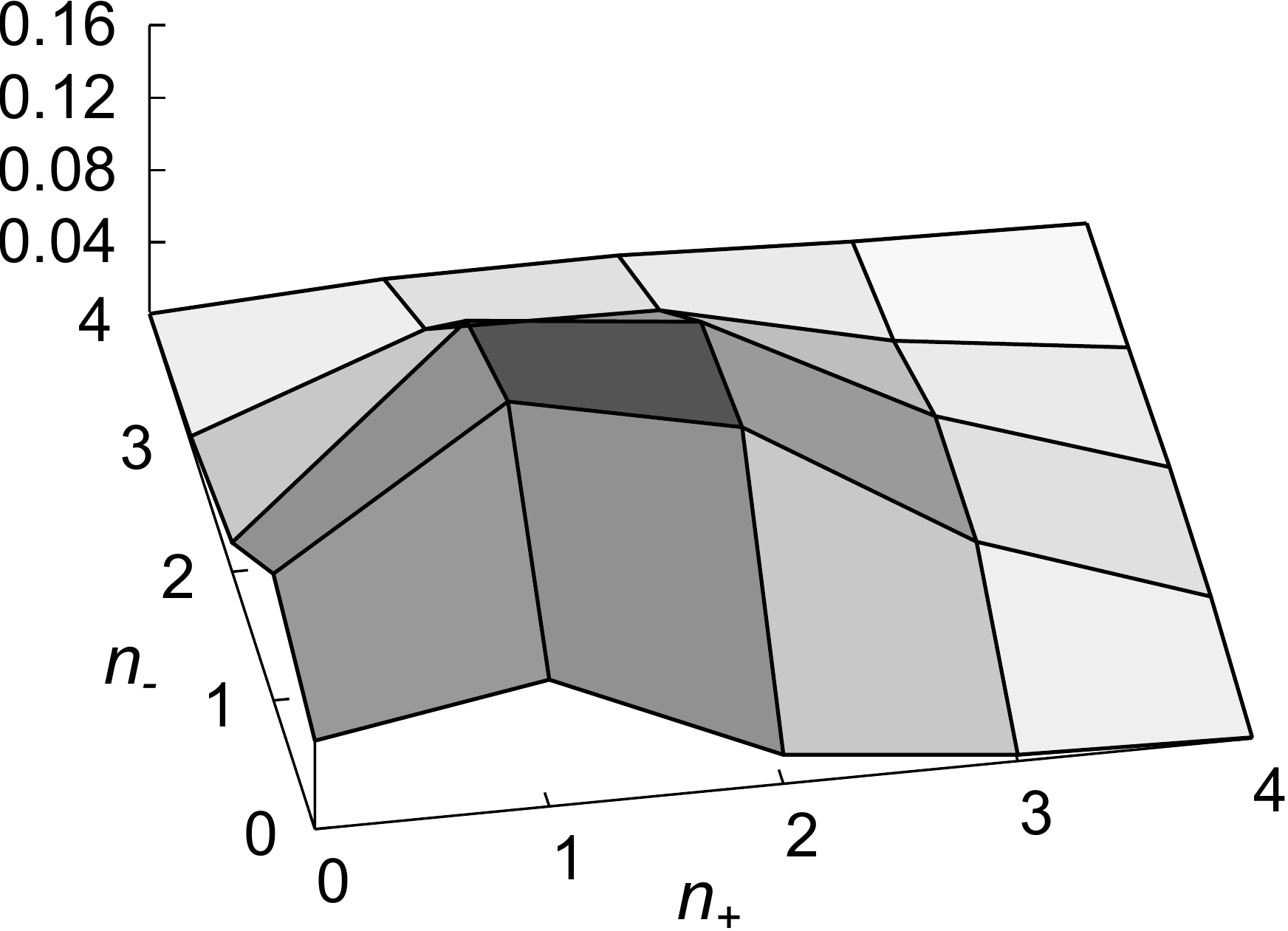}
\end{minipage}\begin{minipage}[hbt]{0.25\textwidth}
\begin{center}\textbf{(d)}\end{center}
\includegraphics[width=0.8\textwidth]{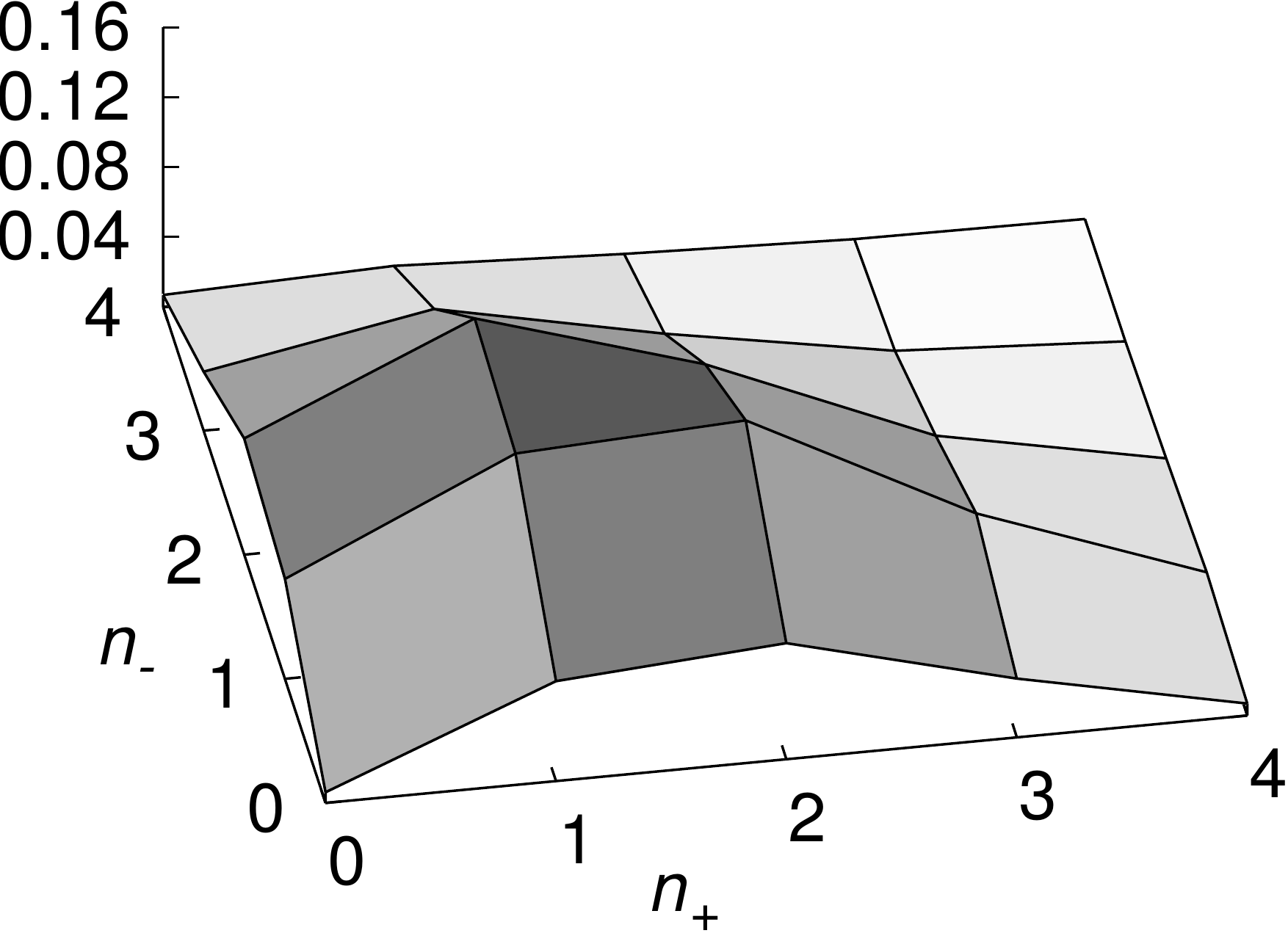}
\end{minipage}
\caption{The probability $p(n_+,n_-)$ of $n_+$ "$+$"-motors and $n_-$ "$-$"-motors bound to the filament \textbf{(a)} in the MF-model and \textbf{(b)} for the EPB-model in the case of no activation. The probability $\tilde{p}(n_+,n_-)$ of $+$"-motors and $n_-$ "$-$"-motors which are actually engaged in the tug-of-war (i.e. which apply a non-zero force on the cargo) in the EPB-model is shown without activation in \textbf{(c)} and with a mutual activation with $a=5$ and $R_A = 32$ nm in \textbf{(d)}.}
\label{mot_4}
\end{figure}

In this first subsection we consider no mutual motor activation. With the
chosen set of parameters given in TABLE~\ref{parameter} the MF-model produces
fast cargo motion with a bimodal velocity distribution around $\pm v_f$,
corresponding to a bimodal histogram of attached motors
(FIG.~\ref{mot_4}\textbf{(a)}).
For the EPB-model, however, 
these high motility states are not observed (FIG.~\ref{mot_4}\textbf{(b)}).
This result gives strong evidence that
fluctuations of the motor positions
play a crucial role in the cargo dynamics,
and that the MF assumption of an equal sharing of forces
between motors of the same kind qualitatively changes the results. 

For the EPB-model, the distribution of bound motors $p(n_+,n_-)$ in
FIG.~\ref{mot_4}\textbf{(b)} has a peaked structure on the diagonal, i.e. most
of the time the same number of motors of both teams are bound to the filament.
Furthermore, in the frame of the EPB-model, we can ask how many of those bound motors are actually engaged in the
tug-of-war, i.e. apply a non-zero force to the cargo. This quantity
$\tilde{p}(n_+,n_-)$ shown in FIG.~\ref{mot_4}\textbf{(c)} demonstrates that
not all motors which are bound to the filament exert a force to the cargo, a
fact which, again, contradicts the mean-field assumption.

\begin{figure}
\centering
\includegraphics[width=0.4\textwidth]{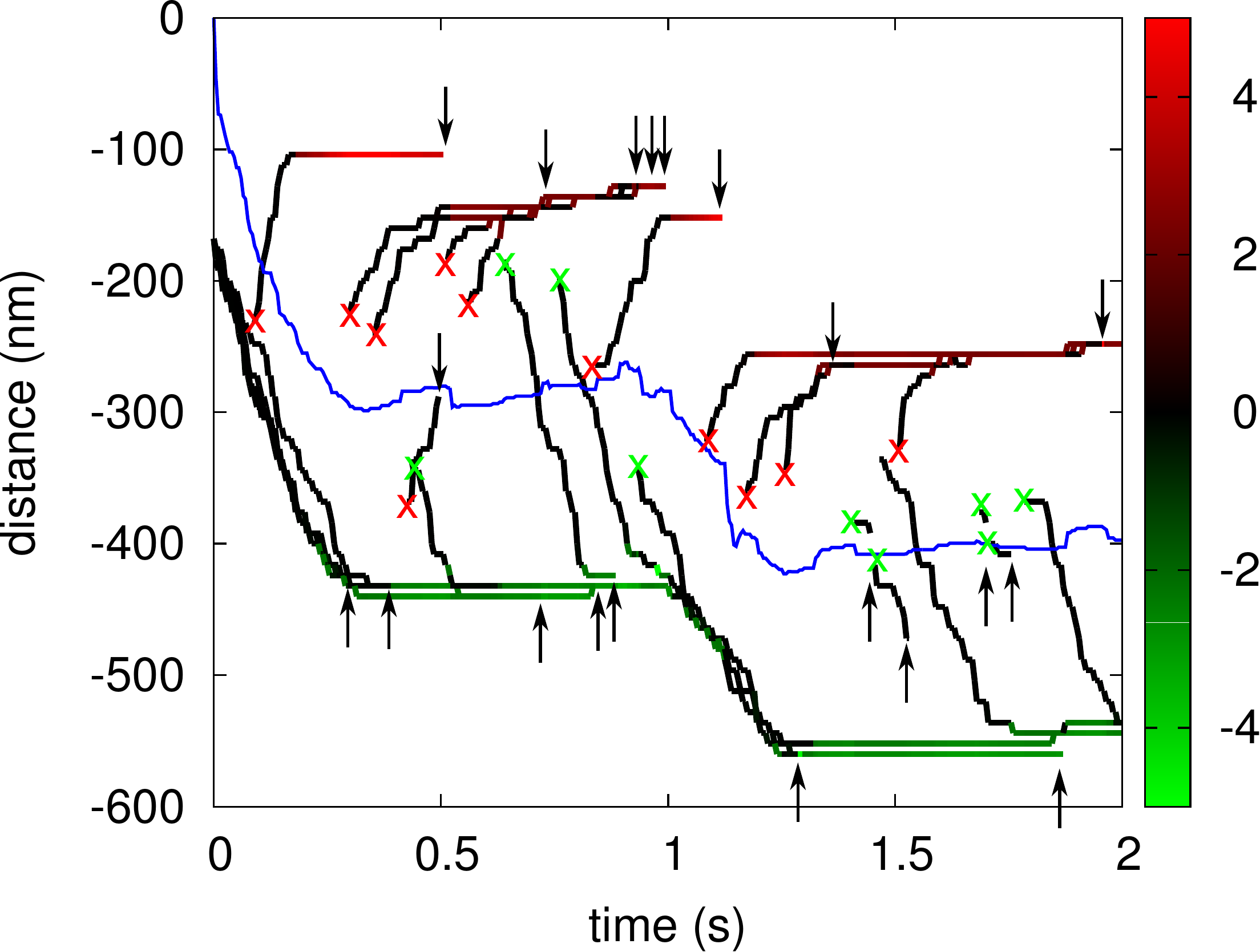}
\caption{One specific realization of the trajectories of the cargo (blue) and of all "$+$"- (red) and "$-$"-
(green) motors. The arrows indicate a detachment event and the crosses an
attachment of a "$+$"- (red) or a "$-$"- (green) motor. The color code gives
the ratio $k_d(F_i)/s_\pm(F_i)$. The initial state is chosen
such that it would correspond to a high motility state in the MF-model,
with no "$+$"-motor attached, and four "$-$"-motors
pulling the cargo. We see here that this state is not stable.}
\label{fig_example}
\end{figure}

When a motor detaches due to a high load, this results
in the MF-model into a new cargo velocity and a sharing of the high load
between the remaining motors of the team, which most probably will also detach.
In the EPB-model, motors have different positions.
As illustrated in FIG.~\ref{fig_example},
the motor that detaches is most probably the most distant from the cargo,
with a reservoir of bound motors of the same team behind.
After detachment, the cargo can move backwards, and as a result,
first, the load between both teams is slightly relaxed.
Second, due to the backward motion of the cargo, some bound motors
which were not exerting any force will become involved in the pulling of
the cargo and replace the detached motor.
This effect makes obvious why we cannot detect the detachment cascades
as they are observed within the mean-field description,
and why high motility states can be sustained only in the MF-model.

\subsection{Mutual motor activation}
In the previous discussion we pointed out that, in the relevant
parameter range, the MF-model can exhibit some high motility states
(i.e. bimodal distributions of the velocity) that are not present
in the EPB-model\footnote{We exclude the cases where only one motor at a
time is pulling, for example in the case of a detachment rate much higher
than the attachment rate. In these trivial cases no collective effect occurs
anyhow.}. This is because the MF-assumption implicitly introduces
some synchronization between the motors.
We shall now explicitely model some mutual motor activation in the EPB-model,
and explore whether this allows to recover a double-peaked histogram of pulling
motors as obtained in the MF-model.

The mutual motor activation is introduced as follows.
If the
$i$-th motor of one team makes a step, the stepping rate of the motors of the
same team within the interval $[x_i - R_A, x_i + R_A]$ is multiplied by a
factor $a$. Here we have chosen $a=5$ and $R_A=32$ nm, i.e. 4 tubulin
subunits. 

We first analyze the influence of activation for a small number $N=4$ of motors of both teams. FIG.~\ref{mot_4}\textbf{(d)} shows that the activation does not change the distribution of attached motors significantly. The probability to have no motor pulling $\tilde{p}(0,0)$ goes to zero and those for the "two against one"-scenarios $\tilde{p}(2,1)$ and $\tilde{p}(1,2)$, respectively, are slightly increased. However, we cannot detect a double-peaked distribution.
\begin{figure}[bt]
\begin{minipage}[hbt]{0.25\textwidth}\centering\textbf{(A)} \includegraphics[width=\textwidth]{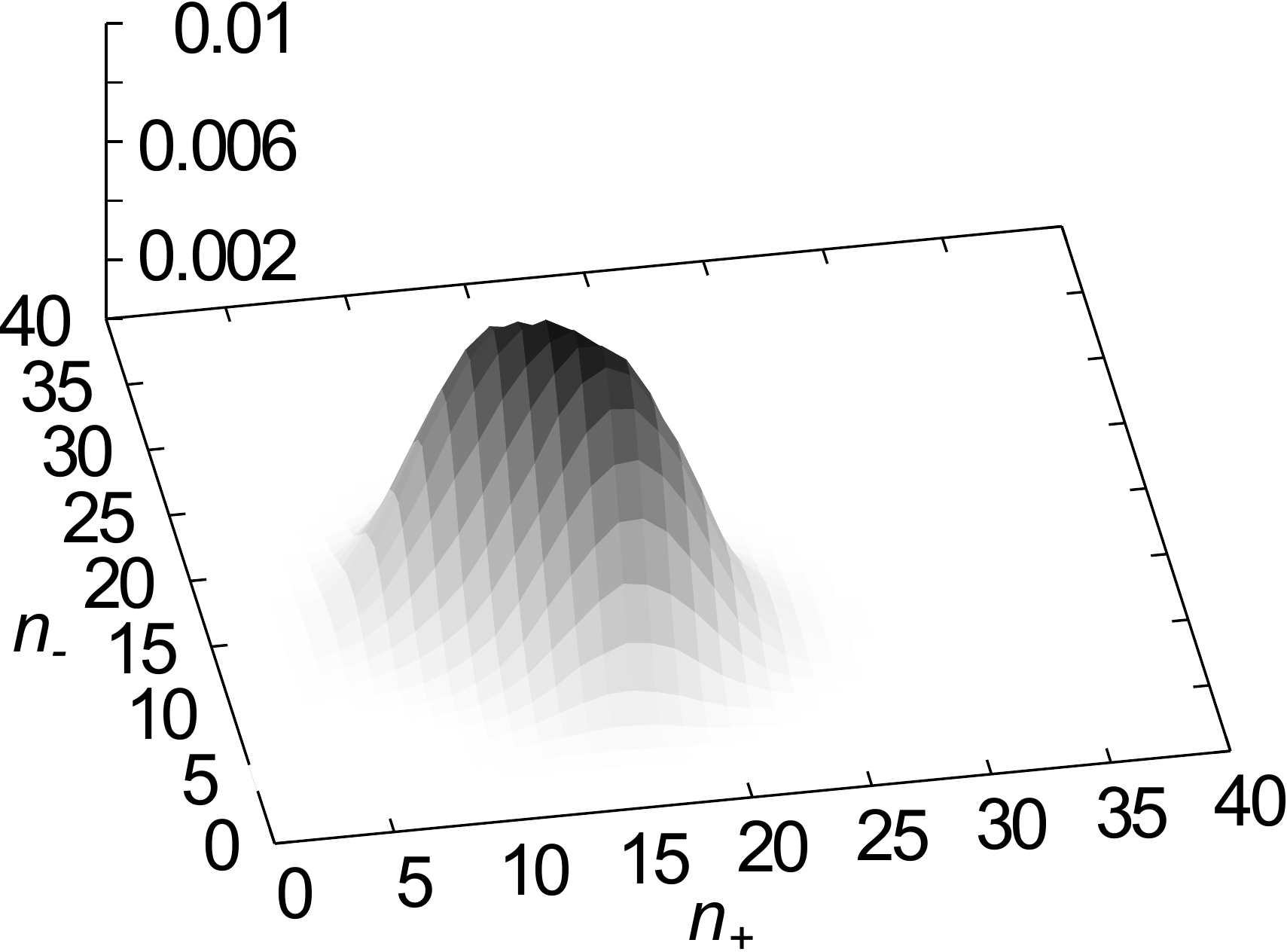}\end{minipage}\begin{minipage}[hbt]{0.25\textwidth}\centering\textbf{(B)} \includegraphics[width=\textwidth]{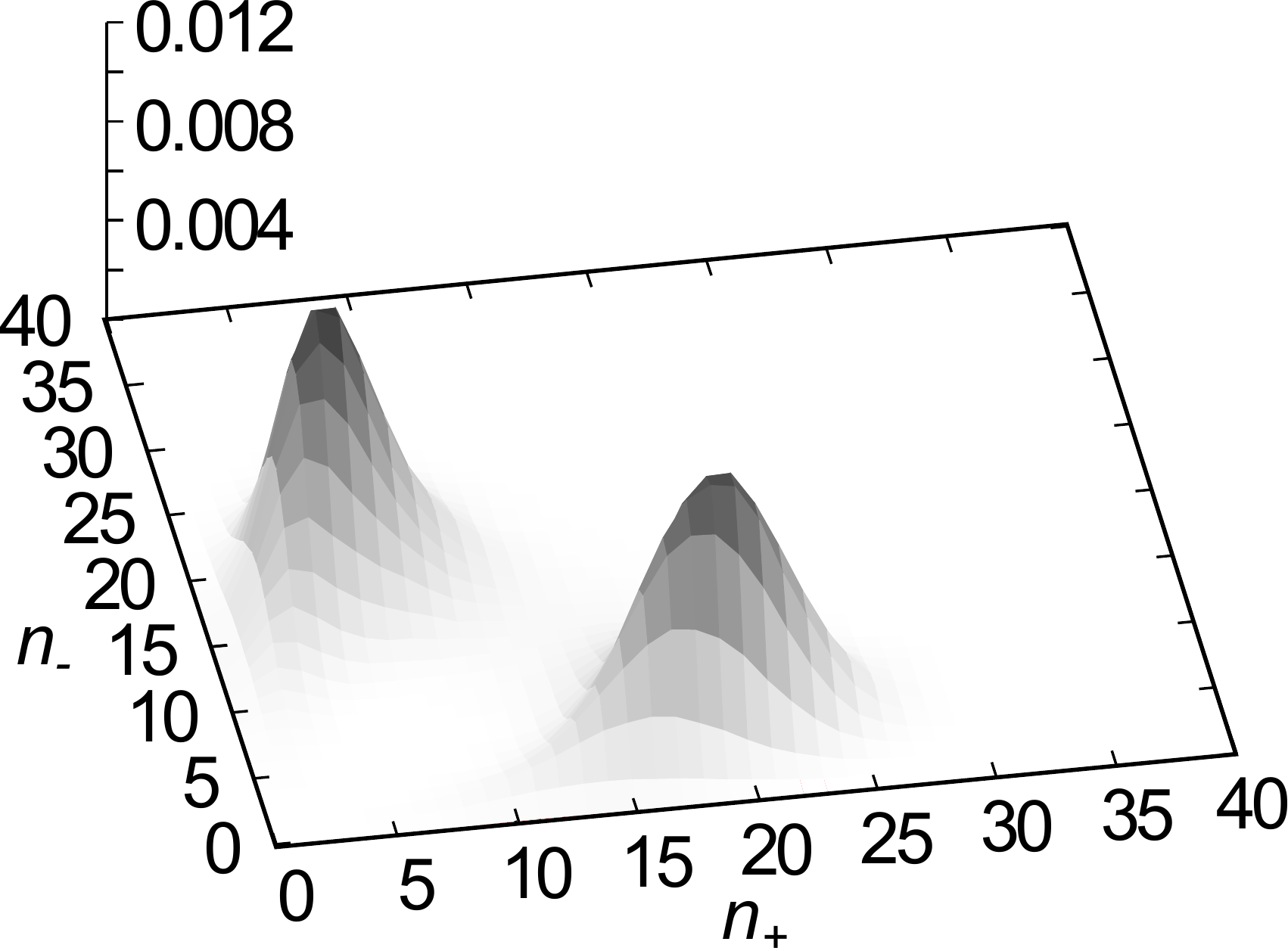} \end{minipage}\\ 
\vspace*{0.2cm}
\begin{minipage}[hbt]{0.25\textwidth}  \includegraphics[width=0.95\textwidth]{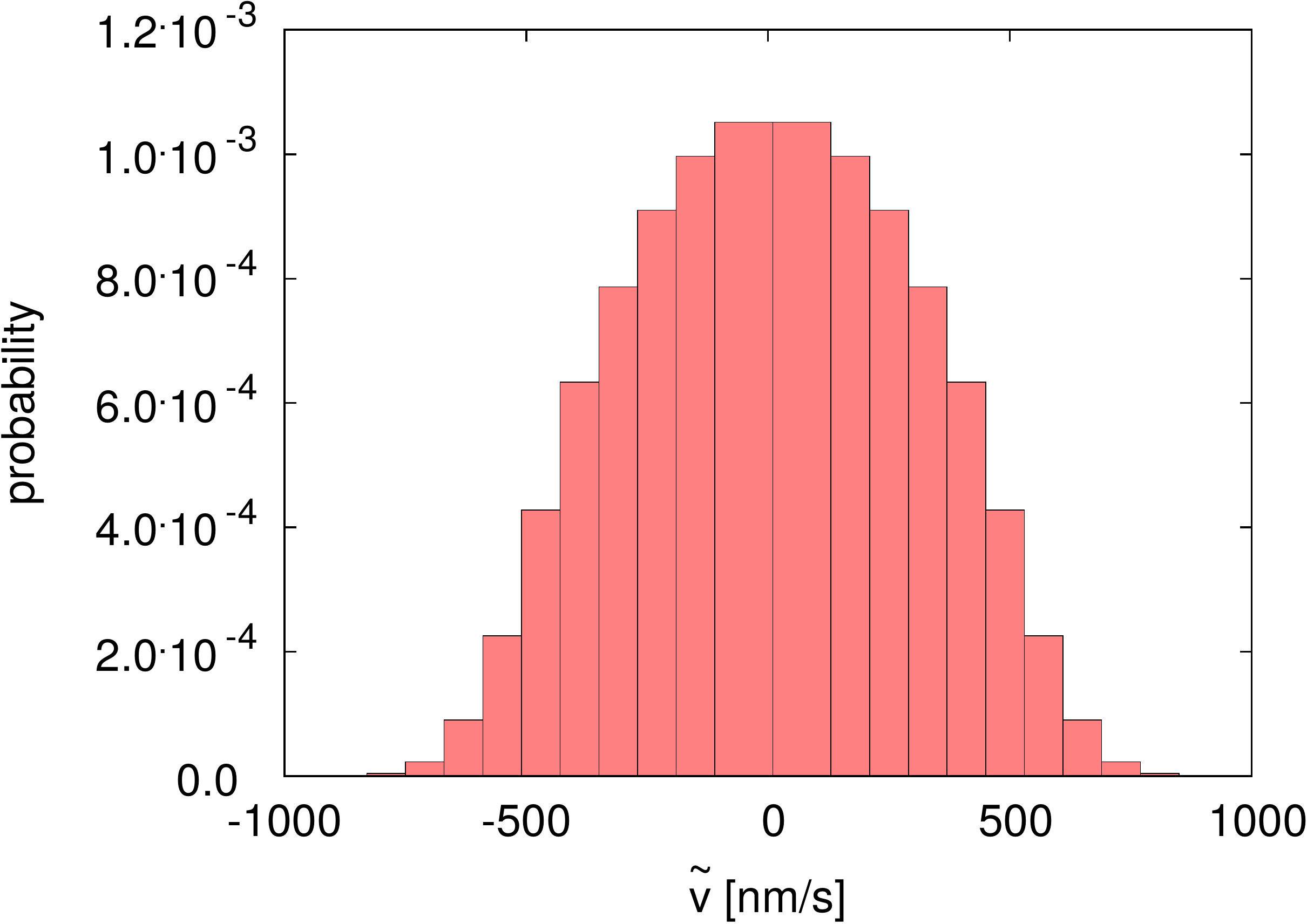}\end{minipage}\begin{minipage}[hbt]{0.25\textwidth} \includegraphics[width=.95\textwidth]{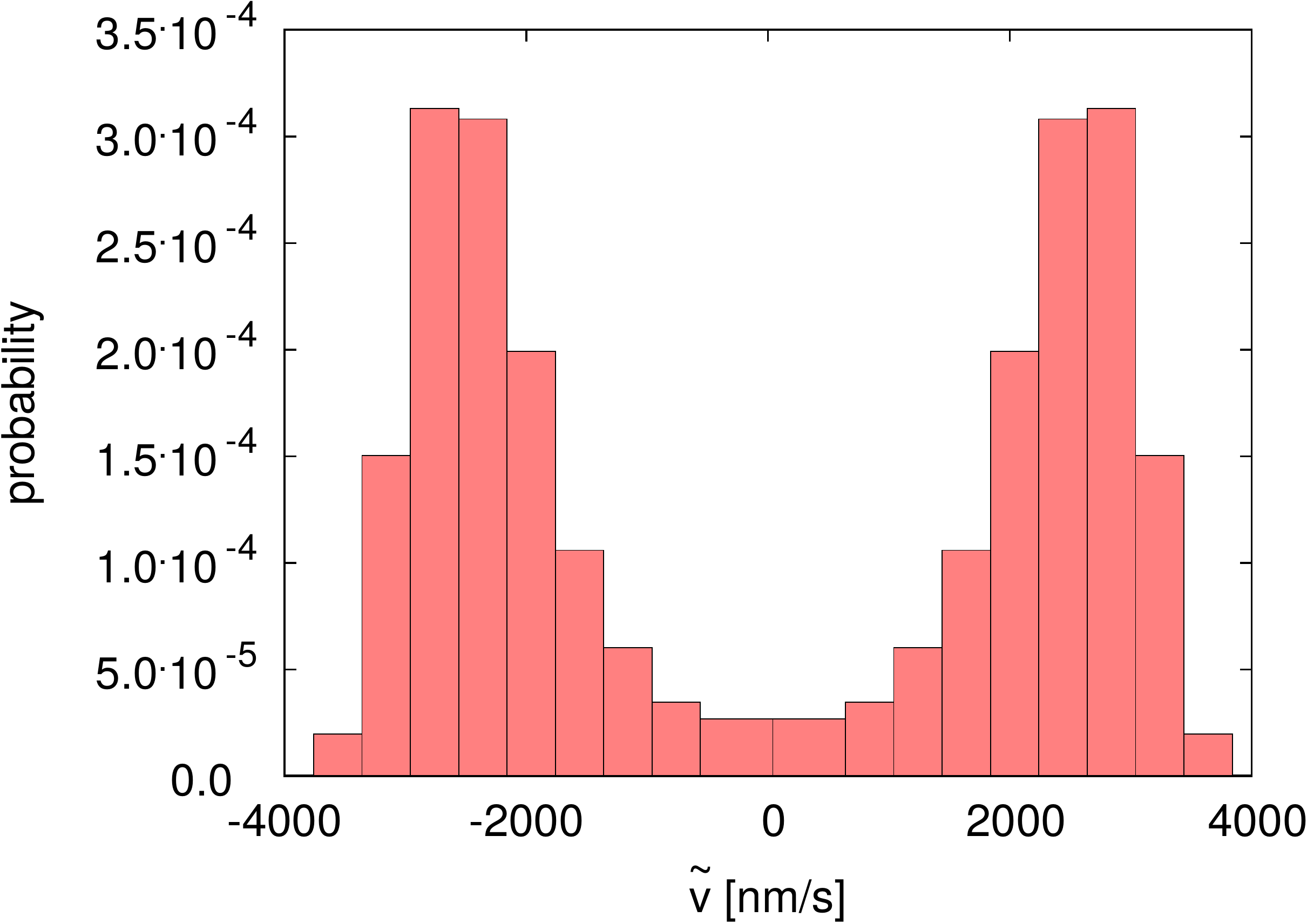}\end{minipage}
\caption{The probabilities $\tilde{p}(n_+,n_-)$ (top) and $p(\overline{v})$
(bottom) are shown for $a=1$, so without activation on the left hand side and
for $a=5$ and $R_A=32$ nm on the right hand side.
}\label{mot_40}
\end{figure}

Now the motor number is increased from $N=4$ to $N=40$ motors per team. In column \textbf{(A)} of FIG.~\ref{mot_40} the probability of engaged motors $\tilde{p}(n_+,n_-)$ (top) and the velocity distribution $p(\overline{v})$ (bottom) without activation ($a=1$) are shown. The velocity distribution is peaked around zero and the maximal probability of engaged motors is still on the diagonal. If we introduce the mutual motor activation we observe the bimodal velocity distribution and probability of engaged motors as shown in column \textbf{(B)} of FIG.~\ref{mot_40}. Apparently, the mean-field assumption is justified for the case of a high number of motors and if, in addition, a synchronization mechanism between the motors exists.
\section{Discussion}

In this work we discuss the
modeling of motor-induced bidirectional cargo
transport inside cells. Apart from the MF-model of M\"uller \etal \cite{mueller_k_l2008} we
also introduce the EPB-model which 
considers explicitly the positions of the motors bound to the filament in the
spirit of \citep{kunwar2011}. This allows us to calculate the forces which act 
on each individual motor explicitly, rather than implicitly
via a mean-field assumption as in the MF-model.
In contrast to the
MF-model, the EPB-model does not exhibit any high
motility state corresponding to a bimodal distribution in velocity and number
of attached motors.

This result is rather robust and we checked that it does not depend on the particular realization of the motor-cargo coupling or the details of the motors response
to external forces. While high motility states are very stable in the MF-model, a realization with only one team pulling the cargo is only a rather rare event with short lifetime in the EPB-model.
In order to  illustrate this further we introduced the mutual activation of molecular motors as a possible mechanism for synchronization of the 
motors' motion. When applied to a large number of motors, this mechanism led indeed to a stabilization of high motility states, comparable to the results of the MF model. 
Therefore, we have shown that fluctuations of motor positions are relevant for motor-cargo system, and that the mean-field assumption corresponds implicitly to an synchronization of the motors which are attached to the cargo.
A spontaneous synchronization only via the tug-of-war between 
the attached motors can be ruled out.  

In view of this result one has to discuss the experimental relevance of high motility states. From our point of view there is no clear evidence that 
these states have actually been observed experimentally,
in spite of the fact that, for example,
the bimodal velocity distributions found in \cite{Hendricks2010}
have been interpreted as an evidence for high motility states.
Indeed, in this work, like in many others, the 
trajectories of bidirectionally transported cargos are not analyzed directly,
but cut into runs and pauses. This decomposition of the trajectories follows 
different definitions in different experiments.
A common problem of this decomposition is that it can induce artifacts in the velocity distribution. For example, applying the criteria by \cite{gross2000} to the trajectories
produced by the EPB-model also gives a bimodal velocity distribution (see Supplementary Material) though this is not found 
if we use the full data set.
More generally, it would be helpful if full trajectory data would be systematically provided from experiments.

Still, several characteristics of cargo transport have been
found experimentally in \vivo experiments~\cite{caspi_g_e2002,kulic2008},
including anomalous diffusion (sub- or superdiffusion
depending on the observation time scales).
We have shown indeed in previous work \cite{EPL14,EPJST14} that,
using more biologically relevant motor characteristics in the EPB-model,
these sub- or superdiffusive particle motion at short times can be reproduced.
Still, \vivo experiments are not so simple to interpret.
\Vitro experiments with purified motors would allow to
have a better control of the system parameters, and to go
even further in the comparison with the models and in the
testing of various scenarios.

\begin{acknowledgments}
This work was supported by the Deutsche 
Forschungsgemeinschaft (DFG) within the collaborative 
research center SFB 1027 and the research training group GRK 1276.
\end{acknowledgments}
\bibliographystyle{apsrev4-1}
\bibliography{biblio.bib}

\clearpage
\section{Supplementary Material}
One quantity to characterize intracellular transport is the velocity with which
a cargo moves. In experiments it is rather difficult to define this velocity.
It depends in particular on the measurement time window,
and there is no common procedure to determine it. One popular method is to cut
trajectories into runs and pauses. Runs are usually defined by giving thresholds concerning their minimal length, minimal 
duration and/or minimal velocity
\cite{gross2000,Hendricks2010}.
Due to the different ways to measure velocity, it is rather hard to compare data from different teams. 

\begin{figure}[htb]
\centering
\includegraphics[scale = 0.3 ]{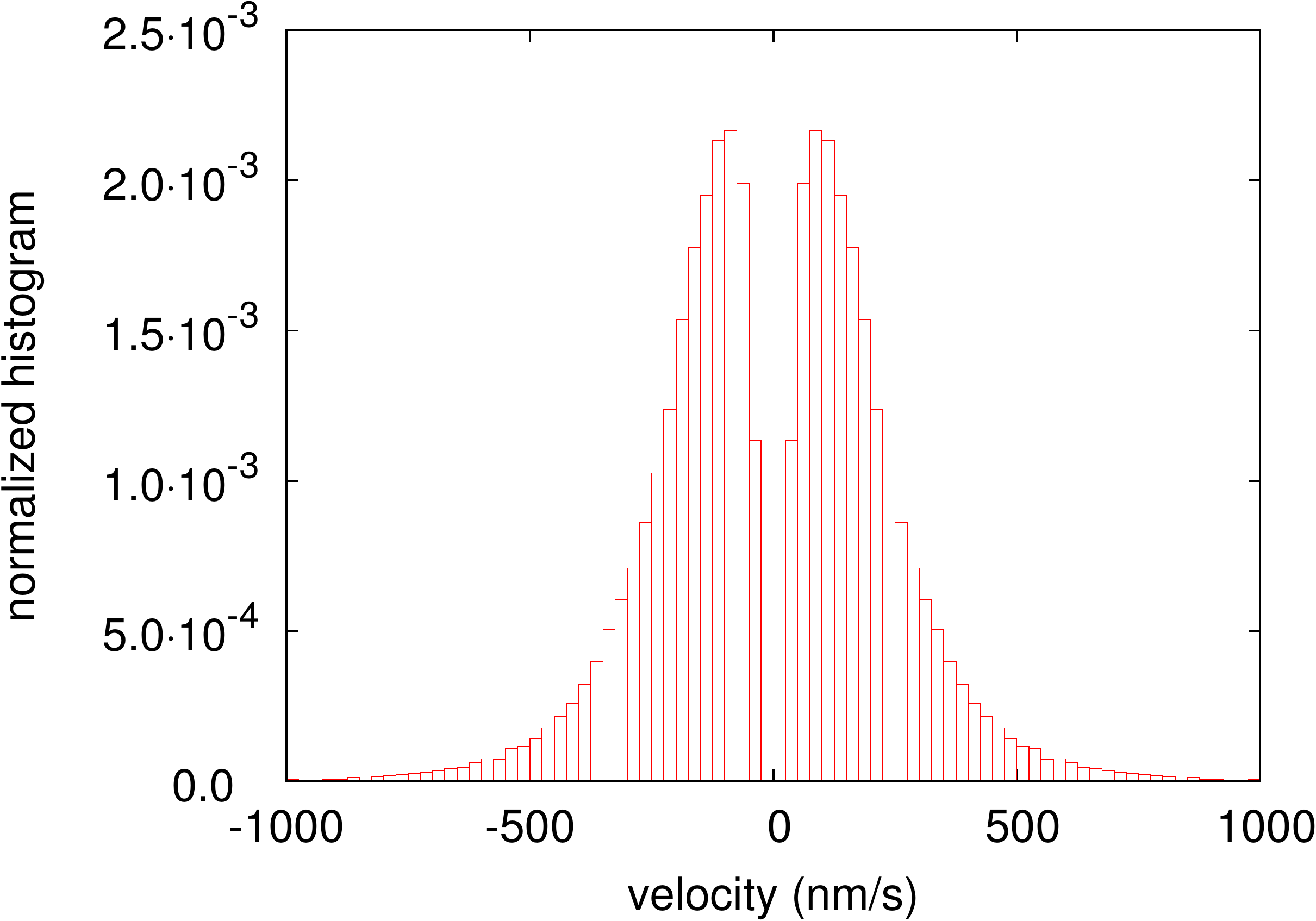}
\caption{Velocity distribution of the EPB-model if we use the method of \cite{gross2000} to determine runs and pauses.
Here we took the parameters of TABLE~\ref{parameter} and $N=4$.}
\label{velo_bimodal}
\end{figure}

Here we want to motivate to use the full data set to characterize cargo motion. We have shown in the main text that the EPB-model produces a velocity distribution which is peaked around zero if we use the whole trajectory to determine it. If we now use the method of \cite{gross2000} we also see a bimodal velocity distribution (FIG. \ref{velo_bimodal}). It is obvious why we find this structure: If we ignore the pauses the probability to find small velocities goes to zero. So by definition the distribution is bimodal. 

From our point of view this kind of determination of velocity distributions by only taking parts of the data into account can lead to misleading results. 

\end{document}